\documentclass[twocolumn,showpacs,preprintnumbers]{revtex4}
\usepackage{graphicx}
\usepackage{dcolumn}
\usepackage{bm}
\usepackage{subfigure}

\begin{document}
\newcommand{\ds}{\displaystyle}
\newcommand{\be}{\begin{equation}}
\newcommand{\en}{\end{equation}}
\newcommand{\bea}{\begin{eqnarray}}
\newcommand{\ena}{\end{eqnarray}}
\title{Static phantom wormholes of finite size}
\author{Mauricio Cataldo}
\altaffiliation{mcataldo@ubiobio.cl} \affiliation{Departamento de
F\'\i sica, Facultad de Ciencias, Universidad del B\'\i o-B\'\i o,
Avenida Collao 1202, Casilla 15-C, Concepci\'on, Chile, and \\
Grupo de Cosmolog\'\i a y Gravitaci\'on-UBB.}
\author{Fabian Orellana}
\altaffiliation{fabian.orellana.verdugo@gmail.com}
\affiliation{Departamento de F\'\i sica, Universidad de
Concepci\'on, Casilla 160-C, Concepci\'on, Chile; and Facultad de
Ingenier\'\i a y Tecnolog\'\i a, Universidad San Sebasti\'an,
Lientur 1457, Concepci\'on 4080871, Chile.
\\}
\date{\today}
\begin{abstract}
In this paper we derive new static phantom traversable wormholes by
assuming a shape function with a quadratic dependence on the radial
coordinate $r$. We mainly focus our study on wormholes sustained by
exotic matter with positive energy density (as seen by any static
observer) and a variable equation of state $p_r/\rho<~-~1$, dubbed
phantom matter. Among phantom wormhole spacetimes extending to
infinity, we show that a quadratic shape function allows us to
construct static spacetimes of finite size, composed by a phantom
wormhole connected to an anisotropic spherically symmetric
distribution of dark energy. The wormhole part of the full spacetime
does not fulfill the dominant energy condition, while the dark
energy part does.

\vspace{0.5cm} 
\end{abstract}
\smallskip
\maketitle 

\section{Introduction}
The accelerated expansion of the universe is one of the most
exciting and significant discoveries in modern cosmology. In the
framework of general relativity, dark energy, which has an equation
of state satisfying the relation $-1<p/\rho<-1/3$, is the most
accepted hypothesis to explain the observed acceleration. However,
there are observational evidences that the cosmic fluid leading to
the acceleration of the universe may satisfy also an equation of
state of the form $p/\rho<-1$, with positive energy density. A
cosmic fluid characterized by such an equation of state is dubbed
phantom energy, and has received increased attention among theorists
in cosmology, since if this type of source dominates the cosmic
expansion, the universe may end in a Big Rip
singularity~\cite{Caldwell} (the positive energy density becomes
infinite in finite time, as well as the pressure). This phantom
energy has a very strong negative pressure and violates the dominant
energy condition (DEC), which can be written as $\rho \geq 0$ and
$-\rho \leq p \leq\rho$. In such a way, late cosmological evidences
cast a serious doubt on the validity of the energy conditions.

Although the cosmic phantom energy is a time dependent matter
source, conceptually it can be also used in study of static
gravitational configurations. An interesting and useful application
is the construction of static wormhole spacetimes, which need to be
sustained by non-standard matter fields, which violates DEC. Note
that the cosmic phantom energy is a homogeneous field with isotropic
pressure. Since, wormholes are inhomogeneous spacetimes, an
extension of phantom energy must be carried out. Specifically, for
static wormholes the phantom matter is considered as an
inhomogeneous and anisotropic fluid, with radial pressure satisfying
the relation $\omega=p_r/\rho <-1$.

In general, in spherically symmetric spacetimes the radial pressure
and the lateral one are different, so one must require the model to
satisfy the DEC specified by $\rho \geq 0$ and $-\rho \leq p_i \leq
\rho$, where $p_i$ are the radial and lateral pressures.

The study of phantom wormholes involves mainly asymptotically flat
phantom wormhole solutions~\cite{Lobo}, which extend from the throat
to infinity. Non asymptotically flat phantom wormholes also have
been studied. In Ref.~\cite{Wang} such wormholes are considered, and
spacetimes extend from the throat to infinity, so they are glued to
the external Schwarzschild solution. Asymptotically and non
asymptotically flat phantom wormholes are also discussed in
Ref.~\cite{Jamil}. Non asymptotically flat phantom wormholes also
have been studied in three dimensions~\cite{Jamil15}. All these
spacetimes also extend to spatial infinity.

In Ref.~\cite{Sushkov} the notion of phantom energy is also extended
to inhomogeneous and anisotropic spherically symmetric spacetimes.
The author finds an exact wormhole solution and shows that a spatial
distribution of the phantom energy is mainly limited to the vicinity
of the wormhole throat.

It is interesting to note that evolving wormholes supported by
phantom energy also have been studied in the presence of a
cosmological constant~\cite{Cataldo15} and without
it~\cite{Cataldo15A}. In both cases the equation of state of the
radial pressure has the form $\omega_r=p_r(t,r)/\rho(t,r)<-1$, with
constant state parameter $\omega_r$ (see also
Ref.~\cite{Cataldo15AA} for a slight generalization of dynamical
phantom equation of state).

This paper presents phantom traversable wormholes by resorting to a
quadratic shape function. For constructing them we use the
conventional approach of Morris and Thorne based on the assumption
of particular forms of the shape function $b(r)$, and the redshift
function $\phi(r)$, in the metric~\cite{Morris:1988cz}
\begin{eqnarray}\label{general spherically BH metric}
ds^2=e^{2 \phi(r)} dt^2- \frac{dr^2}{1-\frac{b(r)}{r}}-r^2
\left(d\theta^2+\sin^2\theta d\varphi^2 \right).
\end{eqnarray}
We assume that the matter source threading the wormhole is described
by a single anisotropic fluid characterized by $T_\mu
^\nu=diag(\rho,-p_r,-p_l,-p_l)$. Therefore, the Einstein field
equations are given by
\begin{eqnarray}
\kappa \rho(r)=\frac{b^{\prime}}{r^2}, \label{rho} \\
\kappa p_r(r)=2\left(1-\frac{b}{r}  \right)
\frac{\phi^{\prime}}{r}-\frac{b}{r^3}, \label{pr} \\
\kappa p_l(r)=\left( 1-\frac{b}{r} \right) \times \nonumber
\\ \left[\phi^{\prime \prime}+\phi^{\prime \, 2} - \frac{b^\prime
r+b-2r}{2r(r-b)} \, \phi^\prime - \frac{b^\prime
r-b}{2r^2(r-b)}\right], \label{pl}
\end{eqnarray}
where $\rho$ is the energy density, $p_r$ and $p_l$ are the radial
and lateral pressures respectively, and the prime denotes the
derivative $d/d r$.

The paper is organized as follows. In Sec. II we study Morris-Thorne
wormholes characterized by a quadratic shape function. In Sec. III
we discuss energy conditions and the positivity of energy density.
In Sec. IV we construct phantom wormholes of finite size. In Sec. V
we conclude with some remarks.



\section{Wormholes with quadratic shape functions}
Now we will study Morris-Thorne wormholes by using a quadratic shape
function in the form
\begin{equation}\label{eq71}
b(r)=a_1r^2+a_2r+a_3
\end{equation}
where $a_1,a_2$ and $a_3$ are constant parameters. In order to have
a wormhole we must impose the Morris-Thorne constraints on the shape
function, so $a_1,a_2$ and $a_3$ are not all free parameters, and
they must satisfy specific constraints which we will now discuss.

First of all, the wormhole must have a minimum radius $r=r_0$, where
the wormhole throat is located. This requirement is expressed
by~\cite{Morris:1988cz,Morris:1988tu}
\begin{eqnarray}\label{r0}
 b(r_0) = r_0,
\end{eqnarray}
and $r_0$ is the minimum value of the radial coordinate $r$. On the
other hand, the shape function must satisfy the condition
\begin{eqnarray}\label{br1}
\frac{b(r)}{r} \leq 1,
\end{eqnarray}
in order to the metric~(\ref{general spherically BH metric}) remains
Lorentzian ($g_{rr} \leq 0$).

Evaluating the shape function~(\ref{eq71}) at the throat, i.e.
imposing the fulfilment requirement~(\ref{r0}), we obtain
\begin{equation}\label{eq72}
b(r)=(r-r_0)\left( a_1 r-\frac{a_3}{r_0} \right)+r,
\end{equation}
and the metric~(\ref{general spherically BH metric}) takes the form
\begin{eqnarray} \label{eq73}
ds^2=e^{2\phi(r)}dt^2-\frac{dr^2}{\left( 1-\frac{r_0}{r}
\right)\left(\frac{a_3}{r_0}-a_1r  \right) }-
\nonumber \\
r^2 \left(d\theta^2+\sin^2\theta d\phi^2 \right).
\end{eqnarray}
It becomes clear that traversable versions of Schwarzschild
wormholes are obtained if $a_1=0$ and $a_3=r_0$.

Now for having a wormhole geometry the shape function must satisfy
the flare-out condition, which is given by~\cite{Morris:1988cz}
\begin{eqnarray*}
\frac{b-b^\prime r}{2b^2} >0.
\end{eqnarray*}
From this equation we obtain
\begin{eqnarray}\label{conditionas}
a_3-a_1 r^2 >0.
\end{eqnarray}
This condition allows us to classify and construct three classes of
wormhole solutions. Namely:

{\bf Case1:}
\begin{eqnarray}
a_1 <0, a_3>0,
\end{eqnarray}
and wormhole exists for $0<r_0 \leq r < \infty $.

{\bf Case 2:}
\begin{eqnarray}
a_1 <0, a_3<0,
\end{eqnarray}
and wormhole exists for
\begin{eqnarray}\label{case2}
\sqrt{\frac{a_3}{a_1}}<r_0< r < \infty.
\end{eqnarray}

{\bf Case 3:}
\begin{eqnarray}
a_1 >0, a_3>0,
\end{eqnarray}
and wormhole exists for
\begin{eqnarray}\label{case3}
0<r_0<r < \sqrt{\frac{a_3}{a_1}}.
\end{eqnarray}

For $a_1>0$ and $a_3<0$ we have static spherically symmetric
gravitational configurations which are not wormholes.

It should be noted that these constraints should be compatible with
the condition~(\ref{br1}), which is expressed in the form $\left(
1-\frac{r_0}{r} \right)\left(\frac{a_3}{r_0}-a_1r \right)>0$, as we
can see from the radial metric component of Eq.~(\ref{eq73}). Since
we have that $r \geq r_0$ for the radial coordinate in a wormhole
geometry, we conclude that also it is necessary to satisfy the
inequation
\begin{eqnarray}\label{15}
\frac{a_3}{r_0}-a_1r >0.
\end{eqnarray}
For the case $1$ this constraint is satisfied automatically. For the
case $2$ we obtain that Eq.~(\ref{15}) implies that
\begin{eqnarray}\label{15b}
r > \frac{a_3}{a_1 r_0},
\end{eqnarray}
while for the case $3$ we obtain that
\begin{eqnarray}\label{15c}
r < \frac{a_3}{a_1 r_0}.
\end{eqnarray}
It is interesting to note that, in principle, we can make that
ranges allowed by the flare-out condition coincide with ranges
imposed by Eq.~(\ref{15}). This can be performed by requiring that
$\sqrt{\frac{a_3}{a_1}}=\frac{a_3}{a_1 r_0}$. Then we have that
\begin{eqnarray}\label{15A}
a_3=a_1 r_0^2.
\end{eqnarray}
If we put this expression into the radial metric component of the
line element~(\ref{eq73}) we obtain that $g_{rr}^{-1}=a_1(r-r_0)^2
r^{-1}$. This implies that the relation~(\ref{15A}) may be applied
only for the case 2, since for the case 3 we have that $a_1>0$ and
the line element~(\ref{eq73}) becomes non-Lorentzian.

\begin{figure}
\includegraphics[scale=0.3]{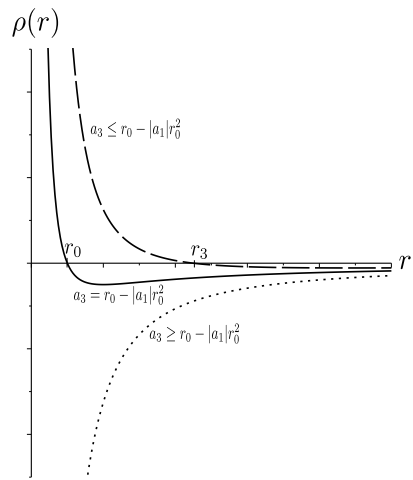}
\caption{The figure shows the qualitative behavior of energy density
for $a_1<0$ and $a_3>0$. For all plots the throat is located at
$r_0$. The dashed line describes the case for which $\rho(r_0)\geq
0$. The solid and dotted lines represent the cases where
$\rho(r_0)=0$ and $\rho(r_0)\leq 0$, respectively. We can see that
in the cases of solid and dotted line the energy density is
everywhere negative, while for the dashed line the energy density is
positive for $r_0 \leq r <r_3$ and becomes negative for
$r>r_3>r_0$.} \label{Fig15AAA}
\end{figure}

\begin{figure}
\includegraphics[scale=0.3]{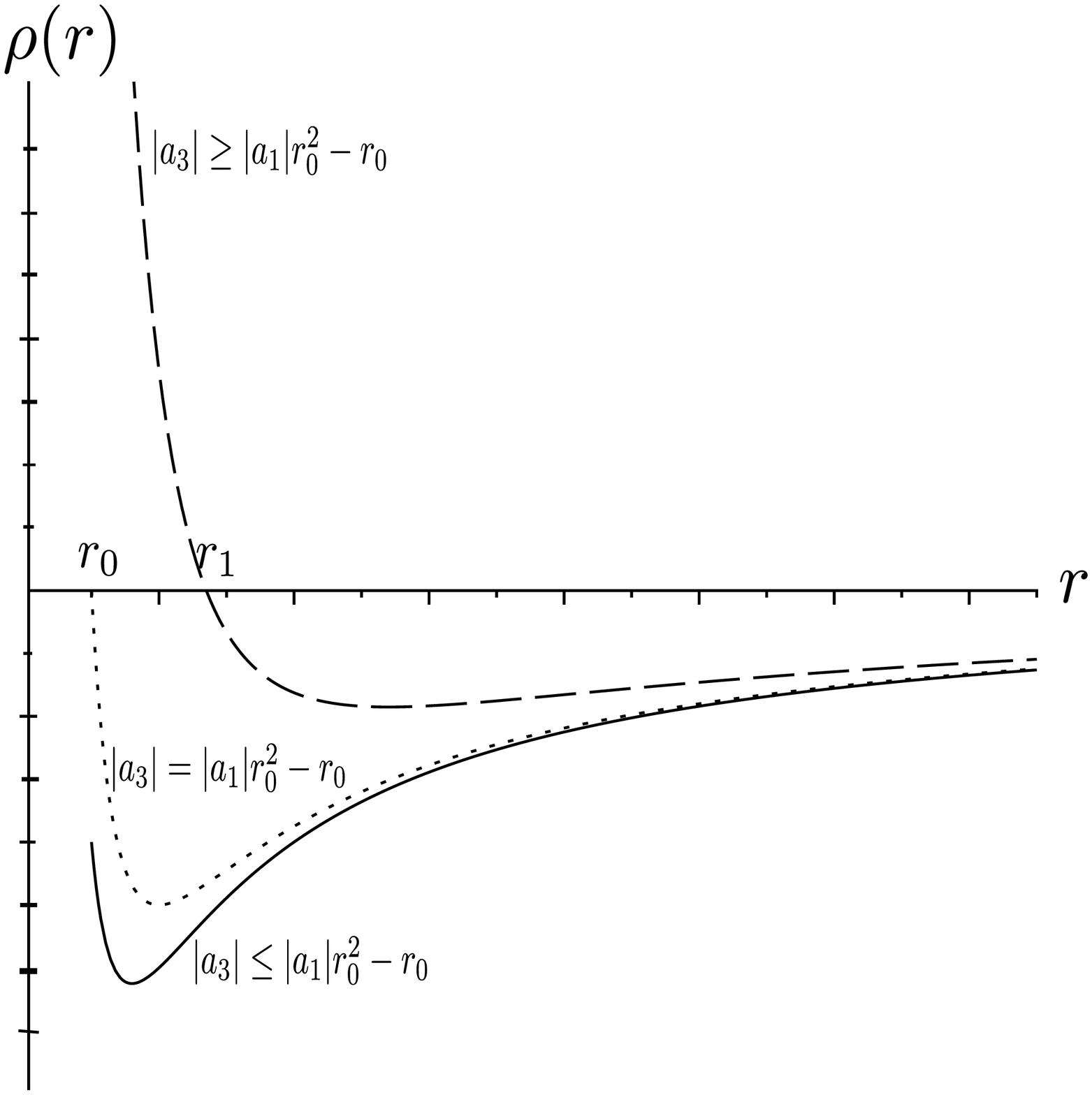}
\caption{The figure shows the qualitative behavior of energy density
for $a_1<0$ and $a_3<0$. The throat is located at $r_0$. The dotted
line describes the case $| a_3 |=r_0^2 | a_1 | -r_0$ for which
$\rho(r_0)=0$. The solid and dashed lines represent the cases $| a_3
| \leq  r_0^2 | a_1 | -r_0$ and $| a_3 | \geq r_0^2 | a_1 | -r_0$,
for which we have at the throat $\rho(r_0)<0$ and $\rho(r_0)>0$,
respectively. We can see that in the case of solid line the energy
density is everywhere negative, while for the dashed line the energy
density is positive for $r_0 \leq r <r_1$ and becomes negative for
$r>r_1$.} \label{Fig15AA}
\end{figure}

\begin{figure}
\includegraphics[scale=0.3]{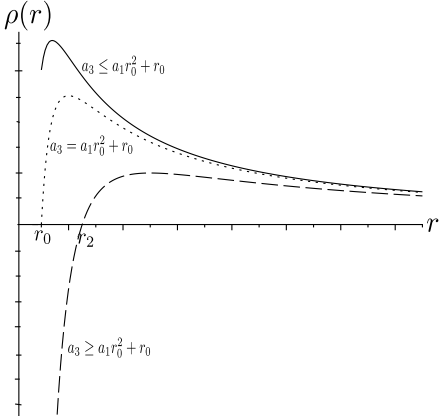}
\caption{The figure shows the qualitative behavior of energy density
for $a_1>0$ and $a_3>0$. The throat is located at $r_0$. The dotted
line represents the case $ a_3 =r_0^2  a_1 +r_0$ for which
$\rho(r_0)=0$. The solid and dashed lines represent the cases $ a_3
 \leq  r_0^2  a_1 + r_0$ and $ a_3  \geq r_0^2  a_1 + r_0$, for
which we have at the throat $\rho(r_0)>0$ and $\rho(r_0)<0$,
respectively. We can see that in the case of solid line the energy
density is everywhere positive, while for the dashed line the energy
density is negative for $r_0 \leq r <r_2$ and becomes positive for
$r>r_2$.} \label{Fig15BB}
\end{figure}

\section{The positivity of energy density and energy conditions}
Since we are interested in finding wormholes supported by phantom
energy, we need to require the positivity of energy density.
Physically, this requirement implies that everywhere any static
observer will measure a positive energy density. Therefore, we shall
study conditions which must satisfy the relevant parameters $a_1$
and $a_3$ in order to have a positive energy density. For the
considered shape function~(\ref{eq72}) the energy density is given
by
\begin{eqnarray}\label{PDE}
\rho=\frac{a_1r_0(2r-r_0)+r_0-a_3}{r_0 r^2}.
\end{eqnarray}

{\bf Case 1:} We consider first the case $a_1<0$ and $a_3>0$ Note
that for large values of radial coordinate we have that $\rho
\approx \frac{2a_1}{r}$, so if  $r \rightarrow \infty$ then $\rho
\rightarrow -0$. The expression~(\ref{PDE}) vanishes at
$r_3=\frac{r_0-a_3+\mid a_1 \mid r_0^2}{2\mid a_1 \mid r_0}$. In
such a way, if $r_3>r_0$, then the energy density is positive for
$r_0 < r<r_3$, while $\rho \leq 0$ for  $r \leq r_3$. The energy
density is everywhere negative for $r \geq r_0$ if $a_3> r_0-\mid
a_1 \mid r_0^2$, and vanishes at $r_0$ if $a_3 = r_0-\mid a_1 \mid
r_0^2$.

{\bf Case 2:} Now, Eq.~(\ref{PDE}) implies that if $a_1<0$ and
$a_3<0$ we have two possibilities to be considered: if
\begin{eqnarray}
\mid a_3 \mid \leq  r_0^2 \mid a_1 \mid -r_0
\end{eqnarray}
then $\rho(r_0) \leq 0$ and the energy density is negative
everywhere for $r > r_0$, while if
\begin{eqnarray}\label{Cambio1}
\mid a_3 \mid >  r_0^2 \mid a_1 \mid -r_0
\end{eqnarray}
then $\rho(r_0) > 0$, and we obtain $\rho(r) > 0$ for $r_0 < r <
r_1$, and $\rho(r) \leq 0$ for $r \geq r_1$, where $r_1=\frac{\mid
a_3 \mid +r_0 +r_0^2 \mid a_1 \mid}{2r_0 \mid a_1 \mid}$.

{\bf Case 3:} Lastly, for $a_1>0$ and $a_3>0$ we have also two
possibilities to be considered: if
\begin{eqnarray}
a_3 \leq  r_0^2 a_1 +r_0
\end{eqnarray}
then $\rho(r_0) \geq 0$ and the energy density $\rho(r) > 0$ for $r
> r_0$, while if
\begin{eqnarray}\label{Cambio2}
a_3 >  r_0^2 a_1 +r_0
\end{eqnarray}
then $\rho(r_0) <0$ and we obtain $\rho(r) < 0$ for $r_0 < r < r_2
$, and $\rho \geq 0$ for $r \geq r_2 $, where $r_2=\frac{a_3+r_0^2
a_1-r_0}{2a_1r_0}$.

In conclusion, phantom wormholes may be constructed for all cases
$a_1<0$, $a_3>0$; $a_1<0$, $a_3<0$ and $a_1>0$, $a_3>0$. In
Figs.~\ref{Fig15AAA}, \ref{Fig15AA} and~\ref{Fig15BB} we show the
qualitative behavior of the energy density for these cases.

Note that the energy density~(\ref{PDE}) may be rewritten in the
form $\rho=\frac{2 a_1}{r}-\frac{a_3+a_1 r_0^2-r_0}{r_0 r^2}$. From
this expression it becomes clear that for positive $a_1$ and $a_3$
we have always a positive energy density by requiring
\begin{eqnarray}\label{Cparacero}
a_3+a_1 r_0^2-r_0 \leq 0.
\end{eqnarray}
If this inequation is not satisfied then the energy density vanishes
at some $r$, changing its sign, as described by Eq.~(\ref{Cambio2}).


Now some words about the energy conditions. It is well known that
the violation of the DEC
\begin{eqnarray}
\rho \geq 0, \rho+p_r \geq 0, \rho+p_l \geq 0
\end{eqnarray}
is a necessary condition for a static wormhole to exist. It is
interesting to note that for the Morris-Thorne metric~(\ref{general
spherically BH metric}), if $\phi(r)=const$, the strong energy
condition $\rho+p_{_{total}} \geq 0$ is satisfied, since the
relation $\rho + p_r + 2p_l =0$ is everywhere valid.

In order to discuss DEC, for simplicity, we shall consider the
zero-tidal-force wormhole version of these quadratic wormholes (i.e.
$\phi(r)=const$). For a such wormhole the energy density is defined
by Eq.~(\ref{PDE}), and the pressures are given by
\begin{eqnarray}
p_r&=&-\frac{ (r-r_0)\left( a_1 r-\frac{a_3}{r_0} \right)+r}{r^3}, \\
p_l&=&\frac{a_3-a_1r^2}{2r^3}.
\end{eqnarray}
By rewriting the radial pressure as
$p_r=-\frac{a_1}{r}-\frac{a_3}{r^3}+\frac{a_3+a_1 r_0^2-r_0}{r_0
r^2}$ we conclude that for positive $a_1$ and $a_3$, if
Eq.~(\ref{Cparacero}) is fulfilled, the radial pressure is
everywhere negative. The lateral pressure vanishes at
$r=\sqrt{a_3/a_1}$, and $p_l>0$ for $r_0 \leq r < \sqrt{a_3/a_1}$,
while $p_l<0$ for $r \geq \sqrt{a_3/a_1}$.

Let us now consider the behavior of $\rho+p_r$ and $\rho+p_l$. For
the first expression we have that
\begin{eqnarray}\label{rhomaspr}
\rho+p_r=\frac{a_1r^2-a_3}{r^3}.
\end{eqnarray}
At the throat this relation gives $\rho+p_r=\frac{1}{r_0} \left(
a_1-\frac{a_3}{ r_0^2} \right)$. It should be noted that the
expression~(\ref{rhomaspr}) vanishes at $r=\sqrt{a_3/a_1}$. For
negative values of $a_1$ and $a_3$ we obtain that $\rho+p_r \leq 0$
for $r \geq r_0$. Now, for positive $a_1$ and $a_3$, from
Eq.~(\ref{case3}) we have that $\sqrt{a_3/a_1}>r_0$, implying that
$a_1<a_3/r_0^{2}$, and then at the throat $\rho+p_r \leq 0$. Since
the expression~(\ref{rhomaspr}) vanishes at $r=\sqrt{a_3/a_1}$, the
weak energy condition is violated at $r_0 \leq r < \sqrt{a_3/a_1}$
as we should expect. For the range $\sqrt{a_3/a_1} \leq r \leq
a_3/a_1$ we have that $\rho+p_r \geq 0$, so DEC may be fulfilled in
this range.

For the expression $\rho+p_l$ we have that
\begin{eqnarray}\label{rhomaspl}
\rho+p_l=\frac{3a_1}{2r}+\frac{a_3}{2r^3}-\frac{a_3+a_1
r_0^2-r_0}{r_0 r^2},
\end{eqnarray}
and at the throat $\rho+p_l=\frac{a_1 r_0^2-a_3+2r_0}{2r_0^3}$,
which in general can be positive as well as negative. However, for
positive $a_1$ and $a_3$ if Eq.~(\ref{Cparacero}) is fulfilled then
the expression in Eq.~(\ref{rhomaspl}) is everywhere positive.

\section{Phantom wormholes of finite size and their embeddings}
Now let us consider embeddings diagrams of the studied wormholes.
The embedding of two dimensional slices $t=const,
\theta=\frac{\pi}{2}$ of the metric~(\ref{eq73}) is performed by
using the embedding function $z(r)$ in equation
\begin{eqnarray}\label{ELE15A}
\frac{dz(r)}{dr}=\left( \frac{r}{b(r)}-1 \right)^{-1/2},
\end{eqnarray}
which takes the form
\begin{eqnarray}\label{ELE15AF}
\frac{dz(r)}{dr}=\sqrt{\frac{(r-r_0)\left( a_1 r-\frac{a_3}{r_0}
\right)+r}{r \left( 1-\frac{r_0}{r}
\right)\left(\frac{a_3}{r_0}-a_1r  \right)}}.
\end{eqnarray}
It becomes clear that, the embedding exists in a Euclidean space if
the expression under the square root is positive. The
Eqs.~(\ref{15})-(\ref{15c}) imply that the denominator is positive,
so we must require the positivity of the numerator, i.e. $b(r)>0$.
The shape function~(\ref{eq72}) is quadratic in $r$, so the
numerator under the square root in Eq.~(\ref{ELE15AF})  may have two
roots, one root, or no roots. The roots of Eq.~(\ref{eq72}) are
given by
\begin{eqnarray}\label{ELE15AR}
r_{\pm}=\frac{1}{2a_1r_0} \left( a_3+a_1r_0^2-r_0\right) \pm
\frac{\sqrt{\Delta}}{2a_1} ,
\end{eqnarray}
where $\Delta=(a_1 r_0+a_3/r_0-1)^2-4 a_1 a_3$. The existence of
real roots depends on values of $\Delta$, therefore the existence of
a wormhole embedding in the Euclidean space depends on values of
$\Delta$.

In the following we will focus our attention on the study of
wormholes with finite size (case 3), therefore we shall consider
wormholes with positive $a_1$ and $a_3$.

{\bf Case $\Delta=0$:} Let us first consider the case where
$\Delta=0$ (i.e. there exists a unique root, and $b(r) \geq 0$).
This condition gives $a_3=r_0+a_1 r_0^2 \pm 2 r_0 \sqrt{a_1 r_0}$,
obtaining two branches for the root~(\ref{ELE15AR}): $r_+=r_-=r_0
\pm \sqrt{r_0/a_1}$. In this case, the shape function is given by
$b(r)=r_0+(r^2+r_0^2-2rr_0)a_1-2 (r-r_0) \sqrt{r_1a_1}$ and the
wormhole extends from $r_0$ to $r_{max}=\sqrt{\frac{r_0+a_1 r_0^2
\pm 2 r_0 \sqrt{a_1 r_0}}{a_1}}$. Notice that, for the negative
branch we have that
\begin{eqnarray}
r_0<\sqrt{\frac{r_0+a_1 r_0^2 -2 r_0 \sqrt{a_1 r_0}}{a_1}}
\end{eqnarray}
if $a_1<1/(4r_0)$, while for the positive branch
\begin{eqnarray}
r_0<\sqrt{\frac{r_0+a_1 r_0^2 + 2 r_0 \sqrt{a_1 r_0}}{a_1}}
\end{eqnarray}
for any $a_1>0$.

{\bf Case $\Delta<0$:} In this case there are not roots, and $b(r)
>0$. Since $a_1>0$, the requirement $\Delta<0$ implies that the embedding always exist
if the parameter $a_3$ satisfies the constraints
\begin{eqnarray}\label{roots}
r_0 \left(1+a_1r_0- 2 \sqrt{a_1r_0} \right)<a_3<r_0 \left(1+a_1r_0+
2 \sqrt{a_1r_0} \right). \nonumber \\
\end{eqnarray}
In other words, if positive $a_1$ and $a_3$ satisfy
Eq.~(\ref{roots}), then the obtained finite size wormhole can be
entirely embedded into the Euclidean space: i.e. the wormhole
spacetime, and simultaneously its embedding in a Euclidean space,
extend from $r_0$ to $r_{max}=\frac{a_3}{a_1r_0}$.

{\bf Case $\Delta>0$:} If condition~(\ref{roots}) is not satisfied,
then roots $r_+$ and $r_-$ of Eq.~(\ref{ELE15AR}) are real, and due
to the positivity of $a_1$, the shape function is positive in the
intervals $(-\infty,r-)\cup (r_+,+\infty)$. Therefore, the embedding
of a constructed finite size wormhole may partially exists in a
Euclidean space. Specifically, an equatorial slice
$\theta=\frac{\pi}{2}$ can be embedded into the Euclidean space in
those ranges obtained from the intersection of intervals
$(-\infty,r-)\cup (r_+,+\infty)$ with the extension of the wormhole
spacetime $\left[r_0,\frac{a_3}{a_1r_0}\right]$.

\subsection{Constructing wormholes}
For constructing explicit examples of phantom wormholes of finite
size we shall use the condition~(\ref{Cparacero}) discussed above.
As we have shown, for positive $a_1$ and $a_3$ the fulfilment of the
condition~(\ref{Cparacero}) ensures that the energy density and
$\rho+p_l$ are everywhere positive, while the radial pressure $p_r$
is everywhere negative. By locating the throat at $r_0=1$ the
condition~(\ref{Cparacero}) becomes $a_1+a_3 \leq 1$. For
simplicity, we shall use the equality $a_1+a_3= 1$.

\begin{figure}
\includegraphics[scale=0.3]{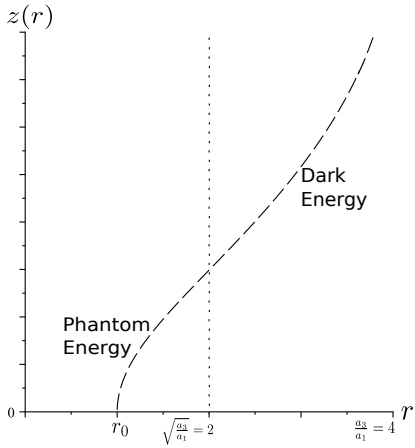}
\caption{The figure depicts the embedding of the
spacetime~(\ref{WH15}). The embedding function is given by
$z(r)=\int \sqrt{\frac{r^2+4}{5r-r^2-4}} \, dr$, and its second
derivative vanishes at $r=2$. The whole spacetime is finite and
extends from $r_0=1$ to $r=4$. The phantom wormhole configuration
extends from the throat to $r=2$, where an inflection point is
present. From $r=2$ to $r=4$ a dark energy distribution is located,
to which is connected the phantom wormhole. For $r<2$ DEC is not
satisfied, while for $r \geq 2$ it does.} \label{Fig15WH}
\end{figure}

\begin{figure}
\includegraphics[scale=0.3]{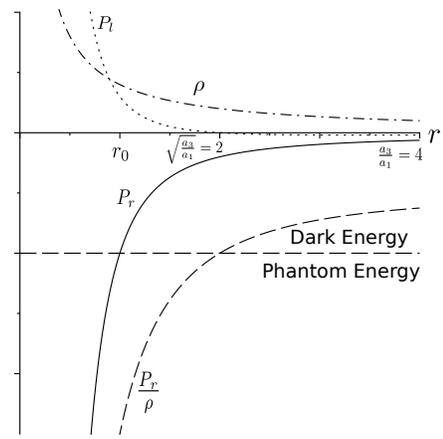}
\caption{The figure compares the behavior of the energy density
(dash-dotted line), radial (solid line) and lateral (dotted line)
pressures; and the equation of state $p_r/\rho$ (dashed line) for
$r_0=1$, $a_1=1/5$ and $a_3=4/5$. Everywhere the energy density is
positive and radial pressure negative; and $p_l
>0$ for $1 \leq r < 2$, and $p_l<0$ for $2 < r \leq 4$. On the other
hand, for $1 \leq r \leq 2$ the equation of state has a phantom
character defined by $-2.5 \leq p_r/\rho \leq -1$, while at $2 \leq
r \leq 4$ the equation of state behaves as dark energy, since $-1
\leq p_r/\rho \leq -0.625$. In such a way, the wormhole is connected
to a distribution of dark energy.} \label{FigFWH15AA}
\end{figure}

\begin{figure}
\includegraphics[scale=0.3]{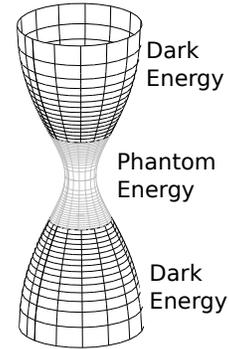}
\caption{The figure shows the three dimensional embedding diagram of
the wormhole of finite size~(\ref{WH15}) with the energy
distribution. } \label{Fig153DWH}
\end{figure}

{\bf Wormhole with positive $\rho$:} Let us first consider the
parameter set $a_1=1/5$ and $a_3=4/5$. From Eq.~(\ref{case3}) we
conclude that the wormhole extends from $r_0=1$ to $r_{max}=2$. The
energy density and pressures are given by

\begin{eqnarray}\label{casoWH1}
\rho=\frac{2}{5r}, p_r=-\frac{r^2+4}{5r^3}, p_l=\frac{4-r^2}{10r^3}.
\end{eqnarray}
It becomes clear that everywhere the energy density is positive and
the radial pressure is negative, while the lateral pressure is
negative for $1 \leq r < 2$ and positive for $r
> 2$. From Eq.~(\ref{casoWH1}) we have that
$\rho+p_r=\frac{r^2-4}{5r^3}$ and $\rho+p_l=\frac{3r^2+4}{10r^2}$,
therefore in the range $1<r<2$ we have that $\rho+p_r<0$ and DEC is
violated. Notice that the metric is given by
\begin{eqnarray}\label{WH15}
ds^2=dt^2-\frac{dr^2}{\left(1-\frac{1}{r}
\right)\left(\frac{4}{5}-\frac{r}{5}\right)}-r^2 \left(d \theta^2
+\sin^2 \theta d \varphi^2 \right). \nonumber \\
\end{eqnarray}
For $r>4$ the spacetime ceases to be Lorentzian, then the spacetime
extends from the throat at $r_0=1$ to $r=4$. So, the whole spacetime
is of finite size, characterized by a wormhole part connected to a
dark energy distribution, which extends from $r=2$ to $r=4$. In this
case the shape function is given by $b(r)=(r-1)(r/5-4/5)+r$, and it
is positive for any value of radial coordinate $r$, implying that
the embedding exists for the whole spacetime~(\ref{WH15}) as shown
in Fig.~\ref{Fig15WH}. Notice that the spacetime in the range $2
\leq r \leq 4$ is supported by a dark energy distribution, which
satisfies DEC. It is interesting to discuss the behavior of the
variable equation of state $p_r/\rho$. For $1 \leq r \leq 2$ we have
the phantom behavior $-2.5 \leq p_r/\rho \leq -1$, while for $2 \leq
r \leq 4$ we have that $-1 \leq p_r/\rho \leq -0.625$, as we would
expect since the wormhole is connected to a distribution of dark
energy (see Figs.~\ref{Fig15WH}-\ref{Fig153DWH}).

{\bf Microscopic wormhole:} It is relevant to note that the wormhole
part of the spacetime can be made arbitrarily small. For doing this
the parameters $a_1>0$ and $a_3>0$ must be chosen in such a way that
$\sqrt{a_3/a_1} \approx r_0$. By using the
condition~(\ref{Cparacero}) we may construct microscopic wormholes
by imposing the equality $a_3+a_1 r_0^2-r_0=0$, implying that the
relations $a_1 \approx \frac{1}{2 r_0}$ and $a_3 \approx
\frac{r_0}{2}$ must be required. As an example, let us consider the
case $r_0=1$. Then we can construct an arbitrarily small wormhole by
making $a_1=\frac{1}{2}-\delta$, $a_3=\frac{1}{2}+\delta$, where
$\delta \approx 0$. Let us put $\delta=0.01$. Then the wormhole
extends from $r_0=1$ to $r=1.0202$, and the whole spacetime to
$r=1.04082$. For energy density and pressures we have that
\begin{eqnarray}
\rho=\frac{0.98}{r}, p_r=-\frac{0.51+0.49 r^2}{r^3},
p_l=\frac{0.51-0.49r^2}{2r^3}. \nonumber  \\
\end{eqnarray}
Clearly, $\rho>0$ and $p_r<0$ everywhere. On the other hand,
$\rho+p_r=\frac{-0.51+0.49 r^2}{r^3}$, $\rho+p_l=\frac{0.51+1.47
r^2}{2r^3}>0$, $-g_{rr}^{-1}=(1-1/r)(0.51-0.49 r) \geq 0$ at $1 \leq
r \leq 1.04082$, and $b(r)=(r-1)(0.49 r-0.51)+r>0$ for $r \geq 1$.
The relation $\rho+p_r$ is negative at $1 \leq r <1.0202$, while
$\rho+p_r \geq 0$ at $1.0202 \leq r \leq 1.04082$. In this case, for
the wormhole part we have that $-1.0204 \leq p_r/\rho \leq -1$, and
for the dark energy distribution part we have $-1\leq p_r/\rho \leq
-0.9804$.

{\bf Wormhole with negative $\rho$:} We can construct also a finite
size wormhole with a negative energy density. In order to do this we
can require that the energy density vanishes at $r=\sqrt{a_3/a_1}$.
This implies that $\frac{a_3+a_1 r_0^2-r_0}{2 r_0
a_1}=\sqrt{\frac{a_3}{a_1}}$. This condition is satisfied by
requiring that $a_1=\frac{a_3+r_0 \pm 2 \sqrt{r_0 a_3}}{r_0^2}$.
This allows us to write the energy density and pressures as (for
positive and negative branches)
\begin{eqnarray}
\rho=-\frac{2(\sqrt{a_3} \pm \sqrt{r_0})(r_0
\sqrt{a_3}-(\sqrt{a_3} \pm \sqrt{r_0})r)}{r_0^2 r^2}, \\
p_r=-\frac{((\sqrt{r_0}\pm\sqrt{a_3})r \mp r_0 \sqrt{a_3})^2}{r_0^2
r^3},
\\
p_l=\frac{-( \sqrt{r_0}\pm\sqrt{a_3})^2r^2+r_0^2 a_3}{2 r_0^2 r^3}.
\end{eqnarray}

These relations give for the radial state parameter
\begin{eqnarray}
\frac{p_r}{\rho}=-\frac{(\sqrt{r_0} \pm \sqrt{a_3})r \mp r_0
\sqrt{a_3}}{2r (\sqrt{r_0} \pm \sqrt{a_3})}
\end{eqnarray}

\begin{figure}
\includegraphics[scale=0.3]{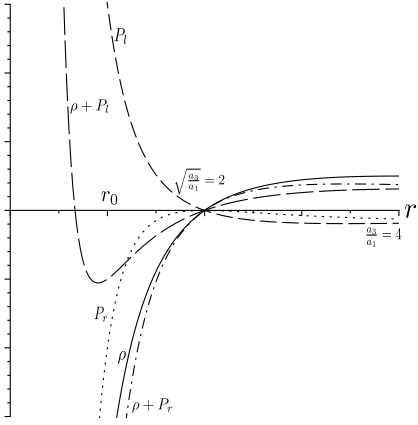}
\caption{The figure compares the behavior of the energy density
(solid line), radial (dotted line) and lateral (dashed line)
pressures, $\rho+p_r$ (dash-dotted line) and $\rho+p_l$ (long-dashed
line) for $r_0=1$, $a_1=1$ and $a_3=4$. It becomes clear that the
energy density is negative for $1 \leq r < 2$ and positive for $2 <
r \leq 4$, while $p_r \leq 0$ for $1 \leq r \leq 4$ and $p_l >0$ for
$1 \leq r < 2$, and $p_l<0$ for $2 < r \leq 4$. On the other hand,
$\rho+p_r$ and $\rho+p_l$ are negative for $1 \leq r < 2$ and
positive for $2 < r \leq 4$.} \label{FigFWH15}
\end{figure}

\begin{figure}
\includegraphics[scale=0.3]{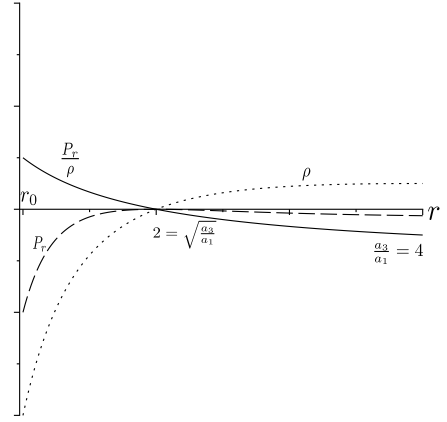}
\caption{The figure compares plots of energy density (dotted line),
radial pressure (dashed line) and the radial equation of state
(solid line) for the case $r_0=1$. $a_1=1$ and $a_3=4$. The energy
density is negative at the wormhole part, while out of it $\rho$ is
positive. In the range $2 \leq r < 4$ we have that $0 \leq p_r/\rho
\leq -1/4$, so DEC is fulfilled.} \label{Fig15rhonegative}
\end{figure}

For an explicit example we take $r_0=1$, $a_1=1$, $a_3=4$. From
expressions of the negative branch we get that
\begin{eqnarray}
\rho=\frac{2r-4}{r^2}, p_r=-\frac{(r-2)^2}{r^3}, p_l=\frac{4-r^2}{2
r^3},
\end{eqnarray}
while $\rho+p_r=\frac{r^2-4}{r^3}$, $\rho+p_l=\frac{(3r-2)(r-2)}{2
R^3}$. The general behavior of these relevant physical magnitudes
are shown in Fig.~\ref{FigFWH15}. It becomes clear that due to
$\rho<0$ at $1 \leq r < 2$, DEC is violated where the wormhole is
located. At $r=2$ this wormhole is connected to an anisotropic
spherically symmetric distribution respecting DEC. In this case for
the radial equation of state we have that $0 \leq p_r/\rho \leq
-1/4$ for $2 \leq r < 4$ (see Fig.~\ref{Fig15rhonegative}).

{\bf Wormhole with mixed energy dependence:} Now we are interested
in construction of finite wormholes with an energy density
exhibiting a mixed dependence, in the sense that the energy density
changes its sign at some sphere of radius $r_0 < r_2 \leq a_3/a_1$,
where $\rho(r_2)=0$, and $r_2=\frac{a_3+r_0^2 a_1-r_0}{2a_1r_0}$. We
have discussed above that if energy density vanishes at some $r$,
then at the throat always $\rho(r_0) <0$. Therefore, if
$r_0<r_2<\sqrt{a_3/a_1}$ we have that $\rho(r)<0$ in this range, and
for $r_2<r< a_3/a_1$ the energy density becomes positive. In this
way, the requirement $r_0<\frac{a_3+r_0^2 a_1-r_0}{2a_1r_0}<
\sqrt{a_3/a_1}$ implies that the parameter $a_1$ satisfies
\begin{eqnarray*}
0  & < &  a_1<\frac{1}{4} \left(1+\sqrt{17}\right)
\end{eqnarray*}
while the parameter $a_3$ the condition
\begin{eqnarray*}
   \frac{-a_1^2+a_1+2}{a_1} & < &  a_3<\frac{-a_1^4+a_1^3+2}{a_1^3}+2
   \sqrt{-\frac{a_1^4-a_1^3-1}{a_1^6}}.
\end{eqnarray*}
For example, we may construct such a wormhole for $r_0=1$, $a_1=1$
and $a_3=3$, obtaining for the relevant quantities
\begin{eqnarray*}
\rho=\frac{2r-3}{r^2}, p_r=\frac{-r^2+3r-3}{r^3},
p_l=\frac{3-r^2}{2r^3},
\end{eqnarray*}
and $\rho+p_r=\frac{r^2-3}{r^3}$,
$\rho+p_l=\frac{3(r^2-2r+1)}{2r^3}$.

The change of sign of the energy density may also occur for a some
radius between $\sqrt{a_3/a_1}$ and $a_3/a_1$. In this case we must
require that $a_3>a_1+2 \sqrt{a_1}+1$ for any positive $a_1$. This
implies that for the wormhole structure the exotic energy density is
always negative, while for the anisotropic spherically symmetric
distribution of matter respecting DEC, the energy density changes
its sign.

Lastly, we may construct finite wormhole solutions for which the
energy density is negative for the whole spacetime, i.e. at $r_0
\leq r \leq a_3/a_1$. By requiring that the energy density vanishes
at $a_3/a_1$ we assure that $\rho(r)<0$ everywhere.  From the
condition $\frac{a_3+r_0^2 a_1-r_0}{2a_1r_0}=\frac{a_3}{a_1}$, we
obtain that if $r_0=\frac{1}{2}$ we must require $a_1=2$ and
$a_3>2$. If the throat is located at $0<r_0<\frac{1}{2}$ we must
require that $0<a_1<\frac{r_0}{r_0^2-2r_0+1}$, while for
$\frac{1}{2}< r_0 \leq a_3/a_1$ (with $r_0 \neq 1$) the condition
$a_1 > \frac{r_0}{r_0^2-2r_0+1}$ must be satisfied. In the last two
cases we have that $a_3=\frac{a_1 r_0^2-r_0}{2r_0-1}$. Notice that
for $r_0=1$ is not possible to construct such a spacetime with
$\rho(a_3/a_1)=0$.

\section{Conclusions}
In this paper we derived new static spherically symmetric
traversable wormholes by assuming a shape function with a quadratic
dependence on the radial coordinate $r$, and it is shown that there
exist wormhole spacetimes sustained by phantom energy. In order to
do this, we specify the equation of state of the radial pressure for
the distribution of the energy density threading the wormhole by
imposing on it a phantom equation of state of the form $p_r/\rho
<-1$. It should be noted that for a quadratic shape function we have
an equation of state $p_r/\rho$ with a variable character. We mainly
focus our study on wormholes sustained by exotic matter with
positive energy density, as seen by any static observer.

An important feature of the wormhole description with a quadratic
shape function is that it includes phantom wormhole spacetimes
extending to infinity, as well as static spacetimes of finite size,
composed by a phantom wormhole connected to an inhomogeneous and
anisotropic spherically symmetric distribution of dark energy. For
latter wormhole types we can construct solutions with phantom matter
confined to a finite region around the throat, which is connected to
the dark energy distribution. The wormhole part does not fulfill the
dominant energy condition, while the dark energy distribution part
does.

Summarizing, in general for finite wormholes ($a_1>0$ and $a_3>0$)
the exotic matter threading the phantom wormhole extends from the
throat at $r_0$ to the sphere of radius $r_{max}=\sqrt{a_3/a_1}$,
and the whole spacetime extends to the square of this $r_{max}$. The
matter source of the gravitational configuration at $\sqrt{a_3/a_1}
\leq r \leq a_3/a_1$ is of dark energy, so it satisfies DEC.

Finally, let us note that in general the spacetime~(\ref{eq73}) is
not asymptotically flat. As a result, the matter distribution for
wormholes extending to infinity must be cut off at some radius
$r=r^*>r_0$ and joined to an exterior asymptotically flat spacetime,
such as, for example, the vacuum Schwarzschild spacetime without
cosmological constant (note that the studied wormholes~(\ref{eq73})
satisfy Einstein equations in the absence of cosmological constant).
On the other hand, for wormholes with finite dimensions, in which
the phantom matter distribution extends from the throat $r_0$ to the
radius $r_{max}=\sqrt{a_3/a_1}$, and the dark energy distribution
extends from $\sqrt{a_3/a_1}$ to $r=a_3/a_1$, the matching to the
exterior vacuum Schwarzschild spacetime can be performed at
$r^*=\sqrt{a_3/a_1}>r_0$ or $r^*=a_3/a_1>r_0$. In other words, the
discussed here wormhole spacetimes can be considered as an interior
solution, which must be matched to an exterior solution, such as the
Schwarzschild geometry, at some radius $r^*>r_0$.

The procedure of construction of traversable wormholes through
matching an interior wormhole solution to the exterior Schwarzschild
solutions is discussed by authors in Ref.~\cite{Wang}. In order to
do this matching one must apply the junction conditions that follow
from the theory of general relativity.

Due to the spherical symmetry of the spacetime, the components
$g_{\theta \theta}$ and $g_{\phi \phi}$ are already
continuous~\cite{Wang}, so one needs to impose continuity only on
the remaining metric components $g_{tt}$ and $g_{rr}$ at $r = r^*$,
i.e.
\begin{eqnarray*}
g^W_{tt}(r^*)=g^{Schw}_{tt}(r^*), \\
g^W_{rr}(r^*)=g^{Schw}_{rr}(r^*).
\end{eqnarray*}
These requirements, in turn, lead to following restrictions for the
redshift and shape functions
\begin{eqnarray*}
\phi^W(r^*)=\phi^{Schw}(r^*), \\
b^W(r^*)=b^{Schw}(r^*).
\end{eqnarray*}
In such a way, the exterior and interior solutions become identical
at the sphere boundary $r = r^*$.

It is interesting to note that for spherically symmetric spacetimes,
one can use directly the field equations to perform the match at the
boundary $r^*$. Einstein equations allow us to determine the energy
density and stresses of the surface $r=r^*$ necessary to have a
match between the interior and exterior spacetimes. If there are no
surface stress-energy terms at the surface $r^*$, the junction is
called a boundary surface. On the other hand, if surface
stress-energy terms are present, the junction is called a thin shell
(see Lemos et al.~\cite{Wang} for a nice review of this issue).

{\bf Acknowledgements:}
This work was supported by Direcci\'on de
Investigaci\'on de la Universidad del B\'\i o-B\'\i o through grants
N$^0$ DIUBB 140708 4/R and N$^0$ GI121407/VBC.


\begin{thebibliography}{2}
\bibitem{Caldwell}
  R.~R.~Caldwell, M.~Kamionkowski and N.~N.~Weinberg,
  Phys.\ Rev.\ Lett.\  {\bf 91}, 071301 (2003)
\bibitem{Lobo} R.~Lukmanova, A.~Khaibullina, R.~Izmailov, A.~Yanbekov, R.~Karimov
and A.~A.~Potapov,
  Indian J.\ Phys.\  {\bf 90}, no. 11, 1319 (2016);  Y.~Heydarzade, N.~Riazi and H.~Moradpour,
  Can.\ J.\ Phys.\  {\bf 93}, no. 12, 1523 (2015); F.~S.~N.~Lobo,
F.~Parsaei and N.~Riazi,
  Phys.\ Rev.\ D {\bf 87}, no. 8, 084030 (2013); J.~A.~Gonzalez,
F.~S.~Guzman, N.~Montelongo-Garcia and T.~Zannias,
  Phys.\ Rev.\ D {\bf 79}, 064027 (2009); F.~S.~N.~Lobo,
  Phys.\ Rev.\ D {\bf 71}, 084011 (2005).
\bibitem{Wang} J.~P.~S.~Lemos,
F.~S.~N.~Lobo and S.~Quinet de Oliveira,
  Phys.\ Rev.\ D {\bf 68}, 064004 (2003);
  D.~Wang and X.~H.~Meng,
  Eur.\ Phys.\ J.\ C {\bf 76}, no. 3, 171 (2016);
P.~K.~F.~Kuhfittig,
  Gen.\ Rel.\ Grav.\  {\bf 41}, 1485 (2009).
\bibitem{Jamil}
  M.~Jamil, P.~K.~F.~Kuhfittig, F.~Rahaman and S.~A.~Rakib,
  Eur.\ Phys.\ J.\ C {\bf 67}, 513 (2010).
\bibitem{Jamil15}
  M.~Jamil and M.~U.~Farooq,
  Int.\ J.\ Theor.\ Phys.\  {\bf 49}, 835 (2010).
\bibitem{Sushkov}
  S.~V.~Sushkov,
  Phys.\ Rev.\ D {\bf 71}, 043520 (2005).
\bibitem{Cataldo15}   M.~Cataldo and P.~Meza,
  Phys.\ Rev.\ D {\bf 87}, no. 6, 064012 (2013); M.~Cataldo and S.~del Campo,
  Phys.\ Rev.\ D {\bf 85}, 104010 (2012); M.~Cataldo, P.~Meza and P.~Minning,
  Phys.\ Rev.\ D {\bf 83}, 044050 (2011); M.~Cataldo, S.~del Campo, P.~Minning and P.~Salgado,
  Phys.\ Rev.\ D {\bf 79}, 024005 (2009).
\bibitem{Cataldo15A} M.~Cataldo, P.~Labrana, S.~del Campo, J.~Crisostomo and P.~Salgado,
  Phys.\ Rev.\ D {\bf 78}, 104006 (2008).
\bibitem{Cataldo15AA}  M.~Cataldo, F.~Aróstica and S.~Bahamonde,
  Eur.\ Phys.\ J.\ C {\bf 73}, no. 8, 2517 (2013).
\bibitem{Morris:1988cz}
  M.~S.~Morris and K.~S.~Thorne,
  Am.\ J.\ Phys.\  {\bf 56}, 395 (1988); M.~S.~Morris, K.~S.~Thorne and U.~Yurtsever,
  Phys.\ Rev.\ Lett.\  {\bf 61}, 1446 (1988).%
\bibitem{Morris:1988tu} M. Visser, Lorentzian Wormholes: From Einstein to
Hawking, (AIP, New York, 1995).
%
%
%
%
\bibitem{Mazharimousavi}
S.~H.~Mazharimousavi and M.~Halilsoy,
  Mod.\ Phys.\ Lett.\ A {\bf 31}, no. 34, 1650192 (2016); M.~Jamil and M.~U.~Farooq,
  Int.\ J.\ Theor.\ Phys.\  {\bf 49}, 835 (2010); R.~A.~Konoplya and A.~Zhidenko,
  Phys.\ Rev.\ D {\bf 81}, 124036 (2010).
\bibitem{Sebastian} S.~Bahamonde, M.~Jamil, P.~Pavlovic and M.~Sossich,
  Phys.\ Rev.\ D {\bf 94}, no. 4, 044041 (2016); T.~Bandyopadhyay, U.~Debnath, M.~Jamil, Faiz-ur-Rahman and R.~Myrzakulov,
  Int.\ J.\ Theor.\ Phys.\  {\bf 54}, no. 6, 1750 (2015); S.~Bhattacharya and S.~Chakraborty,
  ``Evolving Wormholes in a viable $f(R)$ Gravity formulation,''
  arXiv:1506.03968 [gr-qc];  M.~Jamil and M.~Akbar,
  arXiv:0911.2556 [hep-th]; M.~U.~Farooq, M.~Akbar and M.~Jamil,
  AIP Conf.\ Proc.\  {\bf 1295}, 176 (2010).
\bibitem{Heydarzade}
  Y.~Heydarzade, N.~Riazi and H.~Moradpour,
  Can.\ J.\ Phys.\  {\bf 93}, no. 12, 1523 (2015).
\bibitem{Kuhfittig} P.~K.~F.~Kuhfittig,
  Am.\ J.\ Phys.\  {\bf 67}, 125 (1999).
%
%
%
%
%
\end{thebibliography}
\end{document}